\documentclass{article}
\usepackage{amsthm,amsmath,amsfonts,amssymb}

\DeclareMathOperator{\clos}{clos}
\DeclareMathOperator{\td}{td}
\newtheorem{theorem}{Theorem}
\newtheorem{lemma}[theorem]{Lemma}
\providecommand{\M}{MSO${}_1$ }
\providecommand{\N}{\mathbb{N}}

\usepackage{graphicx}
\usepackage{ifpdf}
\ifpdf
\fi

\begin{document}

\title{Dynamic Data Structure for Tree-Depth Decomposition}
\author{Zden\v{e}k Dvo\v{r}\'ak \and Martin Kupec \and Vojt\v{e}ch T\r{u}ma${}^1$}
\date{Computer Science Institute, Charles University\\
  Prague, Czech Republic\\
  \texttt{\{rakdver,kupec,voyta\}@iuuk.mff.cuni.cz}.}

\maketitle

\footnotetext[1]{The work leading to this invention has received funding from KONTAKT II LH12095 and SVV 267313.}

\begin{abstract}
We present a dynamic data structure for representing a graph $G$ with tree-depth at most $D$.
Tree-depth is an important graph parameter which arose in the study of sparse graph classes.

The structure allows addition and removal of edges and vertices such that the
resulting graph still has tree-depth at most $D$, in time bounds depending only
on $D$.  A tree-depth decomposition of the graph is maintained explicitly.

This makes the data structure useful for dynamization of static algorithms
for graphs with bounded tree-depth.  As an example application, we give a dynamic data structure for
MSO-property testing, with time bounds for removal depending only on $D$
and constant-time testing of the property, while the time for the initialization and insertion
also depends on the size of the formula expressing the property.
\end{abstract}

The concept of tree-depth, introduced in \cite{nm-treedepth}, appears
prominently in the sparse graph theory and in particular the theory of graph classes
with bounded expansion, developed mainly by Ne\v{s}et\v{r}il and Ossona de
Mendez~\cite{npom-old,grad1,grad2,grad3,npom-nd1,npom-nd2}.  One of its many
equivalent definitions is as follows.  The tree-depth $\td(G)$ of an undirected simple graph $G$
is the smallest integer $t$ for that there exists a rooted forest $T$ of height $t$ with vertex set $V(G)$
such that for every edge $xy$ of $G$, either $x$ is
ancestor of $y$ in $T$ or vice versa---in other words, $G$ is a subgraph of
the closure of $F$.

Alternatively, tree-depth can be defined using (and is related to) rank
function, vertex ranking number, minimum elimination tree or weak-coloring
numbers.  Futhermore, a class of graphs closed on subgraphs has bounded tree-depth if and only if
it does not contain arbitrarily long paths.  Tree-depth is also related to other structural graph
parameters---it is greater or equal to path-width (and thus also tree-width),
and smaller or equal to the smallest vertex cover.

Determining tree-depth of a graph is NP-complete in general.  
Since tree-depth of a graph $G$ is at most $\log(\lvert G\rvert)$ times its
tree-width, tree-depth can be approximated up to $\log^2(\lvert G\rvert)$-factor,
using the approximation algorithm for tree-width~\cite{bodlaender-appx}.
Furthermore, for a fixed integer $t$, the problem of deciding whether $G$ has tree-depth at most $t$
can be solved in time $O(\lvert G\rvert)$.  Minimal
minor/subgraph/induced subgraph obstructions for the class of all graphs of
tree-depth at most $t$ are well characterized, see~\cite{treed-obstructs}.
Clearly, tree-depth is monotone with respect to all these relations.  For more
information about tree-depth, see the book~\cite{da_book}.

A motivation for investigating structural graph parameters such as tree-depth is
that restricted structure often implies efficient algorithms for problems that
are generally intractable.  Structural parameters have a flourishing
relationship with algorithmic meta-theorems, combining graph-theoretical and
structural approach with tools from logic and model theory---see for instance~\cite{grokre}.
A canonical example of a meta-theorem using a structural
parameter is the result of Courcelle~\cite{courcelle} which gives linear-time algorithms
for properties expressible in MSO logic on classes of graphs with bounded
tree-width.

Tree-depth is similar to tree-width,
in the sense that it measures ``tree-likeness'' of a graph and also allows
decomposition with algorithmically exploitable properties.  However, tree-depth
is more restrictive, since bounded tree-depth implies bounded tree-width, but
forbids the presence of long paths.  Long paths turn out to be related to the hardness
of model checking for MSO logic~\cite{lampis-paths,frigro-cplexity}.  This
motivated a search for meta-theorems similar to~\cite{courcelle} on more
restricted classes of graphs, such as the result of Lampis~\cite{lampis-metath}
that provides algorithms with better dependence on the
size of the formula for classes such as those with bounded vertex cover or bounded max-leaf
number. This result was subsequently generalized by Gajarsk\'y and Hlin\v{e}n\'y to graphs with
bounded tree-depth~\cite{gaj-hlin-mso_trees}.

In the usual static setting, the problem is to decide whether a graph given on input has some fixed property $P$.
Our work is of dynamic kind, that is, the considered graph gradually changes over time and we have to be able to answer any time whether it has the property $P$.
One application comes immediately in mind.  Graphs modelling many natural phenomena, such as the web graph, graphs of social networks or graphs of some physical structure all change rapidly.
However, there is another area where this dynamic approach is useful.
For example, one reduces a graph by removing edges, and each time an edge is removed, some procedure has to be performed.
Instead of running the procedure from scratch every time, it makes sense to keep some dynamic information.
Classical examples are the usage of a disjoint-find-union data structure in minimal spanning tree algorithms~\cite{4algs} or Link-cut trees for network flow algorithms~\cite{sle-tarj}.
A more recent example is a data structure for subgraph counting~\cite{dt-sc} with applications in graph coloring and social networking.
 
The main theorem of our paper follows.
\begin{theorem}
Let $\phi$ be a MSO${}_2$ formula and $D\in \mathbb{N}$.
There exists a data structure for representing a graph $G$ with $\td(G)\leq D$ supporting the following operations:
\begin{itemize}
\item insert edge $e$, provided that $\td(G+\{e\})\leq D$,
\item delete edge $e$,
\item query---determine whether $G$ satisfies the formula $\phi$.
\end{itemize}
The time complexity of deletion depends on $D$ only, in particular, it does not depend on $\phi$ or $\lvert G\rvert$.
The time complexity of insertion depends on $\phi$ and $D$, but does not depend on $\lvert G\rvert$.
The time complexity of the initialization of the data structure depends on $\phi$, $D$ and $\lvert G \rvert$.
The query is done in constant time, as is addition or removal of an isolated vertex.
\end{theorem}

The dependence of the initialization and edge insertion is roughly a tower of height $D$ where the highest element of the tower is the number of nested quantifiers of $\phi$ squared.

The basic idea of the data structure is to explicitly maintain a forest of smallest depth whose closure contains $G$,
together with its compact constant-size summary obtained by identifying ``equivalent'' subtrees.  This summary is sufficient
to decide the property expressed by $\phi$, as outlined in the following paragraph.

Two graphs are said to be $n$-equivalent, if they satisfy the same first order formulas with at most $n$ quantifier alterations---that is, for instance, of the form $\forall x_{1..i_1} \exists x_{i_1+1..i_2} \forall x_{i_2+1..i_3} \ldots \exists x_{i_{n-1}+1..i_n} \phi(x_{1..n})$, where $\phi$ is quantifier-free.
This concept of $n$-equivalency is of practical use for model checking.
It serves to reduce the investigated graph to a small one, so that time-expensive approaches as brute force become possible (a technique known as kernelization).
An example of such application is the following theorem (from section 6.7 of~\cite{da_book}): for every $D,n$ exists $N$ such that every graph $G$ with $\td(G)\leq D$ is $n$-equivalent to one of its induced subgraphs of order at most $N$.
This can be extended even to labeled graphs.
In our work we use a similar theorem, taken from~\cite{gaj-hlin-mso_trees}.
Informally, the result says that when one is interested in checking whether a specific formula is true on a class of trees of bounded depth, then one can also assume bounded degree.
This allows us to only maintain the summary of the tree-depth decomposition as described above.

In the rest of the paper, the first section reviews necessary definitions and tools we use, and the second section describes in detail the data structure and its operations.
We conclude the paper with the application to dynamic model checking.

\section{Preliminaries}

In this paper, all trees we work with are rooted.
For simplicity, we assume in this section that all graphs we work with are connected.
If we encounter a disconnected graph, we consider each of its connected components individually.

Let $T$ be a tree, the \emph{depth} of $T$ is the maximum length of a path from the root of $T$ to a leaf of $T$.
Two trees are isomorphic if there exists a graph-isomorphism between them such that the root is preserved under it.
Mostly we will work with trees with vertices labelled from some set of $l$ labels -- two $l$-labelled trees are $l$-isomorphic if they are isomorphic as trees and the isomorphism preserves labels.

The \emph{closure} $\clos(T)$ is the graph obtained from $T$ by adding all edges $(x,y)$ such that $x$ is an ancestor of $y$, and $x\neq y$.
For instance, the closure of a path is a complete graph.
The \emph{tree-depth} $\td(G)$ is the minimum number $t$ such that there exists a forest $T$ of depth $t$ such that $G\subseteq \clos(T)$.
For instance, the tree-depth of a path on $n$ vertices is $\lceil \log_2 (n+1) \rceil$.
A \emph{limb} of a vertex $v\in T$ is the subgraph induced by some of the children of $v$.
A second-order logic formula $\phi$ is in \M logic, if all second-order quantifiers are over sets of elements (vertices) and the language contains just the relation $edge(u,v)$.

The following result is a simplification of Lemma 3.1 from \cite{gaj-hlin-mso_trees}.
\begin{lemma}\label{lem-gh}
Let $\phi$ be an \M sentence, $l,\ D\in \N$.
Then there exists a number $S$ with the following property.
Let $T$ be an $l$-labelled tree of depth at most $D$ with vertices labelled with $l$ labels, and $v$ a vertex of $T$.
If $v$ has more than $S$ pairwise $l$-isomorphic limbs, then for the tree $T'$ obtained by deleting one of those limbs we have that
$$T' \mbox{ satisfies } \phi \iff T \mbox{ satisfies } \phi.$$
\end{lemma}

The Lemma implies in particular that with respect to $\phi$-checking there are only finitely many $l$-labelled trees of depth at most $D$ -- that is, every $l$-labelled tree of depth at most $D$ is $\phi$-equivalent to some $l$-labelled tree of depth at most $D$ and maximum degree at most $S$.
We call such trees $\phi$-\emph{minimal}.

Let $G$ be a graph of tree-depth $D$, the \emph{tree decomposition} $T$ of $G$ is a $2^{D-1}$-labelled tree such that $G\subseteq \clos(T)$, where a vertex $v$ is labelled by a 0-1 vector of length $D-1$ that encodes the edges between $v$ and the vertices on the path from $v$ to the root (1 whenever the edge is present, 0 otherwise).
Let $l_D$ be a set of labels we describe later, \emph{compressed tree decomposition} of the graph $G$ is an $l_D$-labelled tree $C$ obtained from a tree decomposition $T$ of $G$ as follows.
For every vertex, all its limbs that are pairwise-isomorphic are deleted except for one representative, in which we additionally store the number of these limbs.
Vertices of $C$ are called \emph{cabinets}, and the underlying tree decomposition $T$ is called a \emph{decompression} of $C$.
A set of all vertices corresponding to the same cabinet (that is, inducing $l_D$-isomorphic limbs) and having the same vertex as a father in the decompression is called a \emph{drawer}.
Thus every cabinet is disjointly partitioned into drawers.
For an example how a graph, its tree decomposition and compressed tree decomposition look like, see the figures on page \pageref{fig:graph} (in the compressed tree decomposition, the number next to the drawers denotes how many vertices are there in each drawer).

Now we describe the labelling.
We start inductively, with $l_0$ being just a set of vectors of length $D-1$.
Assume that $\phi$ is some fixed formula we specify later (in Section \ref{sec-find_root}) and let $S$ be the number obtained from applying Lemma \ref{lem-gh} to it.
Let $B$ be a cabinet that induces a subtree of depth $t'\leq t$ in $C$.
The label of $B$ consists of the label of a corresponding vertex $b$ of $T$ and of a vector $vec$ with entry for every $l_{t-1}$-labelled $\phi$-minimal tree $M$ of depth smaller than $t'$ with value
$$vec_{M} = \min\{S,\mbox{number of limbs of $b$ which are $\phi$-equivalent to $M$}\}.$$

During the update operations, we will be occasionally forced to have more than one cabinet for a given isomorphism type (that is, a cabinet will have two pairwise-isomorphic limbs).
Both the decomposition and the individual cabinets that have isomorphic children will be called \emph{dirty}.

\begin{figure}
\centering
\begin{minipage}{.5\textwidth}
    \centering
    \includegraphics{example.1}
    \caption{Graph}
    \label{fig:graph}
\end{minipage}%
\begin{minipage}{.5\textwidth}
    \centering
    \includegraphics{example.2}
    \caption{Tree-depth decomposition}
    \label{fig:decomp}
\end{minipage}
\end{figure}
\begin{figure}
    \centering
    \includegraphics{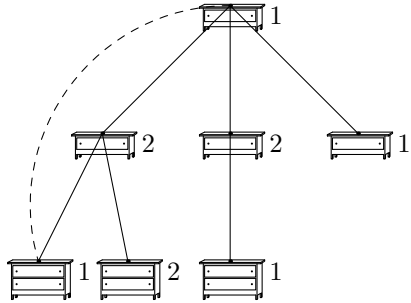}
    \caption{Compressed graph}
    \label{fig:compress}
\end{figure}

\section{Data structure}

Our data structure basically consists of storing some extra information for every vertex $v$ of the represented graph $G$,
and of a compressed tree-depth decomposition $T$ of $G$ with depth at most $D$.
We will store the following for every $v\in G$.
\begin{itemize}
\item Label of the cabinet corresponding to $v$, that is, the vector of its neighbors on the path to root and the vector with the numbers of limbs of $v$ isomorphic to individual $\phi$-minimal trees.
\item Pointer to the father of $v$ in $T$ (more precisely, pointer to a vertex $u$ of $G$ that is the father of $v$ in the decompression of $T$).
However, in some operations we need to change the father of all vertices in a drawer at once -- thus instead of storing father individually for every vertex, for every drawer we will maintain a pointer to the common father of the vertices in this drawer, and every vertex in the drawer will have a pointer to this pointer.
\item Linked list of sons of $v$ in $T$.
This is again implemented by having a linked list of drawers at $v$, and for every drawer in it, a linked list of vertices in this drawer.
\end{itemize}

Additionally, we keep the vertex $v$ which is the root of $T$ and we call it $r$ -- we again assume connectedness of $G$ in this section, otherwise we keep a list of roots corresponding to individual components.

\subsection{Extraction of a path}
In this subsection we describe an auxiliary operation of extracting a path.
It can be seen as a temporary decompression of a part of $T$ in order to make some vertex accessible.
The result of extracting $v$ from $T$ is a dirty compressed tree decomposition $T'$ of $G$, such that on the cabinets in $T'$ corresponding to $r-v$ path there are no cabinets to which corresponds more than one vertex of $G$.

First, we find the vertices of the $r-v$ path, and the corresponding cabinets in $T$. This is done by simply following the father-pointers from $v$, and then by going backwards from $r$, always picking the cabinet that corresponds to the label of the vertex on the $r-v$ path.
Then, for every cabinet $B$ on this path with more than one vertex, let $b$ be the vertex of the $r-v$ path lying in $B$, and $c$ its father -- which we assume to be the only vertex in its cabinet, $C$.
We remove $b$ from the lists of sons of $c$ of the label of $b$, and move $b$ into a new list for $c$, and do the corresponding change in $T'$, that is, creating a new cabinet of the same label as a son of $C$, thus making $C$ a dirty cabinet.

The complexity of this operation is clearly linear in $D$.

\subsection{Edge deletion}
Edge deletion is simple -- let $vu$ be the edge to be deleted, with $v$ the lower vertex (in the tree-order imposed by $T$).
We extract the vertex $v$ from $T$.
Now $u$ lies on the $r-v$ path, and as there are no other vertices in the cabinets on the corresponding path in $T$, we remove the edge $vu$ from the graph and change the labels for the cabinets and vertices accordingly. 
The only affected labels are on the $r-v$ path, and we will precompute during initialization what the label should change into.
It can also happen that removal of such edge disconnects the graph -- this also depends only on labels and thus will be precomputed in advance.
When such situation occurs, we split $T$ into two components -- the new root depends only on the labels, and the vertices for which labels change are only on the $r-v$ path.

Now, we need to clean the dirty cabinets.
As the only dirty cabinets are on the $r-v$ path, we traverse this path, starting from $v$ and going upwards, and for every vertex $w$ in a dirty cabinet, we compare the label of $w$ with the labels of other present drawers at the father of $w$,
and move $w$ to the correct drawer/cabinet.

The complexity of this operation is clearly linear in $D$.

\subsection{Rerooting}
Rerooting is also an auxiliary operation, which will allow us to easily handle the edge insertion.
This operation takes a compressed tree decomposition $T$ and a vertex $r_N$ of $G$, for which we have a guarantee that there is a tree decomposition with depth at most $D$ such that $r_N$ is its root, outputs one such compressed tree decomposition $T'$ and updates data for vertices in $G$ accordingly.
In this subsection we denote by $r_O$ the root of $T$, that is, the old root.

We proceed as follows:
\begin{enumerate}
\item extract $r_N$ from $T$,
\item remove $r_N$ from $T$ entirely,
\item consider the connected components thereof -- those that do not contain $r_O$ have depth $<D$ and thus can be directly attached under $r_N$. Recurse into the component with $r_O$.
\end{enumerate}
Only the third point deserves further explanation.
The components are determined by the labels only, so we will precompute which labels are in which components and what vertices are the roots of the components.
Every connected component of $T-r_N$ that does not contain $r_O$ must have as its highest vertex (under the tree-order) a son of $r_N$, thus these components are already in their proper place.
For the component $C$ with $r_O$, either it has depth $<D$ and thus can be attached under $r_N$, with $r_O$ being a son of $r_N$.
We have to deal with two details -- firstly, there might be some edges to $r_N$ from vertices that were above $r_N$ in $T$ -- but none of these vertices was in a cabinet with more than one vertex, thus we only change the labels accordingly.

Secondly, the limbs of $r_N$ in $T$ that are in $C$ have no father after removal of $r_N$.
But as they are in $C$, for every such limb there is an edge from it to some vertex on the $r_N-r_O$ path $T$.
Choose lowest such vertex, and make it new father for that limb.
This refathering is done by using the pointers for the drawers -- note that every cabinet that is a root of such limb consist only of single drawer, thanks to the extraction of $r_N$.
Thus the total number of operations we have to do is linear in $D$ and the maximum number of children of $r_N$, which is $l_d$.
As in the case of edge deletion, we have to clean dirty cabinets (which are in $C$) in the end.
This can again be done by simply comparing labels on that former $r_N-r_O$ path.

However, it might happen that $C$ has depth exactly $D$.
But we are guaranteed that there exist a tree decomposition with $r_N$ as a root, which implies that there exists a tree decomposition of $C$ with depth $D-1$.
If we know which vertex can serve as a root of such decomposition, we can apply the operation recursively.
We describe the procedure to find a root in Section \ref{sec-find_root}.
An additional thing we have to care about is that some vertices of $C$ have an edge to $r_N$ -- this information has to be preserved in the recursive call.
But the number of such vertices is bounded by a function of $D$ and thus it is not a problem -- we only modify their labels accordingly.
After this recursive call, we again clean dirty cabinets.

The complexity of this operation for one call is linear in $D+l_d+$ time to find new root, and there are at most $D$ recursive calls.

\subsection{Edge insertion}
Let $u,v$ be two vertices not connected with an edge, such that $G+uv$ has treedepth at most $D$, we now describe how to add such edge.
If the edge $uv$ respects the tree-order (that is, either $u$ lies on $v-r$ path or vice versa), we just extract the lower of the two vertices, add the edge, and get rid of dirty cabinets.

Otherwise, there exists a vertex $r_1$ which is a root of some tree decomposition of $G+uv$. We describe the procedure for finding it in Section \ref{sec-find_root}.
Reroot into this vertex to obtain decomposition $T_1$.
Now, $u$ and $v$ must be in the same connected component $C_1$ of $T_1-r_1$.
Again, unless the edge $uv$ respects the tree-order now, we can find a vertex $r_2$ in $C_1$ which is a root of some tree decomposition of $C_1+uv$ of depth at most $D-1$, and reroot into it to obtain decomposition $T_2$ of $C_1$, with $u$ and $v$ lying in the same component $C_2$ of $C_1 - r_2$.
Carrying on in the obvious manner, this process stops after at most $D$ iterations.

The complexity of this operation is $O(D \cdot $ complexity of rerooting$)$.

Finally, we just remark that addition and removal of a vertex (without incident edges) is implemented trivially by just adding/removing new component with the corresponding label.

\subsection{Finding a root} \label{sec-find_root}
Let us recall what we have to face in this section.
We want to find a vertex $v$ such that there is a tree $T$ of depth at most $t'\leq t$ such that its closure contains the connected component $C$ of the graph $G+(a,b)-\{v_1,v_2,\ldots,v_k\}$ as a subgraph and $v$ is a root of $T$.
The vertices $v_1,v_2,\ldots,v_k$ correspond to the roots found in previous applications of this procedure, $(a,b)$ denotes the edge we are trying to add.

At this point we define the formula $\phi$ according to which we constructed the labelling of our trees.
Let $\gamma(C)$ be a formula which is true whenever $C$ is connected --- this is easily seen to be expressible in \M logic ---
and $\tau_d(G)$ the following formula:
$$\tau_d(G) = (\exists v\in G)(\forall C\subseteq G)(\gamma(C-\{v\}) \Rightarrow \tau_{d-1}(C-\{v\})),$$
with $\tau_1(v)$ being always true.
Then $\tau_d(G)$ says that there exists a tree $T$ with depth at most $d$ such that $G\subseteq \clos(T)$.
Furthermore, as we need to express the addition of an edge, we work with the logic with two extra constants $a,b$, and modify the formula for $\gamma$ accordingly to obtain $\gamma'$ and $\tau'$.
The resulting formula $\tau'_t$ is the formula $\phi$.

Using Lemma \ref{lem-gh}, we construct all $\phi$-minimal trees -- note that we have to consider every possible evaluation of the constants $a,b$, that is,
we construct all trees of depth at most $t$ such that no vertex has more than $S$ pairwise-isomorphic limbs, and then for every two of labels, we choose two arbitrary vertices having that label, and choose them to be $a$ and $b$.
For every such minimal tree, we evaluate the formula, that is, we find which vertex is to be the root of the tree decomposition.
It might happen that the formula is false, that is, no such vertex exists, which means we evaluated $a$ and $b$ so that the graph has tree-depth greater than $d$.
But such evaluation will not occur during the run of the structure --- recall that we restricted the edge additions --- and thus we can safely discard these minimal trees.
Thus for every minimal tree we store the label of the vertices that can be made root, and when applying the rerooting subroutine, we find an arbitrary vertex of this label.
This has complexity at most $D$, because when looking for the given vertex, we follow first pointer from the corresponding linked list of children for a vertex.

This means that the total complexity of the edge insertion is $O(D(D+l_D))$.
Finally, let us remark on the complexity of initialization.
From \cite{gaj-hlin-mso_trees} we conclude that $l_D$, that is, the number of $\phi$-minimal trees, is roughly a tower of 2's of height linear in $D$, to the power $\lvert \phi \rvert ^2$.
The complexity of operations we do for every $\phi$-minimal tree is bounded by a polynomial in $D$ and $l_D$.

\subsection{Dynamic model checking}
We now describe how to modify the structure so that it also allows queries of the form ``does $G$ satisfy the formula $\varphi$'', where $\varphi$ is some fixed MSO formula.
The modifications affect only the Section \ref{sec-find_root}.
Instead of using just the formula $\phi$ to obtain the $\phi$-minimal trees, we apply the Lemma \ref{lem-gh} to the formula $\varphi$ also and in the construction of the minimal trees and the labelling, we use the higher of the two numbers obtained from the Lemma.
Then for every such obtained minimal tree we evaluate whether it satisfies $\varphi$ or not, this time without evaluating the constants $a,b$.


\bibliographystyle{siam}
\bibliography{main}
\end{document}